# Scale-invariant statistics of period in directed earthquake network

Sumiyoshi Abe[1, a] and Norikazu Suzuki[2, b]

[1]Institute of Physics, University of Tsukuba, Ibaraki 305-8571, Japan

[2]College of Science and Technology, Nihon University, Chiba 274-8501, Japan

**Abstract.** A new law regarding structure of the earthquake networks is found. The seismic data taken in California is mapped to a growing directed network. Then, statistics of period in the network, which implies that after how many earthquakes an earthquake returns to the initial location, is studied. It is found that the period distribution obeys a power law, showing the fundamental difficulty of statistical estimate of period.

---

[a] e-mail: suabe@sf6.so-net.ne.jp

[b] e-mail: suzu@phys.ge.cst.nihon-u.ac.jp

Seismicity is a research subject of great interest in view of science of complexity [1]. The scale-free natures of the celebrated Omori law [2] and Gutenberg-Richter law [3] indicate existence of long-time correlation and inappropriateness of distinguishing earthquakes by the values of magnitude, respectively. Also, recent works [4] show that both the spatial distance and the time interval between two successive earthquakes follow definite statistical laws. In addition, there is an interesting study [5], which discusses that a major earthquake can trigger aftershocks that are more than 1000 km far from the epicenter. These imply that event-event correlation in seismicity may be long-ranged and, thus, no windows should be put in analysis both spatially and temporally (i.e., the viewpoint of non-reductionism). Moreover, the aging phenomenon and its associated scaling property have been found [6] in the nonstationary seismic time-series intervals termed the Omori regimes (in which the Omori law for aftershocks holds). Taking into account slow relaxation (i.e., power-law nature of the Omori law) and quenched disorder of the stress distribution, this result shows that there are striking similarities between glassy dynamics and the yet unknown mechanism governing aftershocks.

Now, in recent years, the concept of complex networks [7] has also been attracting much attention as a novel approach to complex systems. This stream has been initiated by the characterizations of small-world networks [8] and scale-free networks [9]. The primary purpose of research in this area is to study the topological and dynamical properties of complex systems described as random graphs. Quite recently, the concept of complex network has been introduced in seismology in [10]. The earthquake network is constructed as follows. The geographical region under consideration is divided into a lot of small cubic cells. A cell is regarded as a vertex if earthquakes with any values of magnitude occurred therein. If two successive earthquakes occur in different cells, the corresponding two vertices are linked by an edge, whereas if two successive events occur in an identical cell, then a loop is attached to that vertex. Based on the assumption, which is supported by the results in [4], that all successive events are indivisibly correlated, each edge/loop is supposed to represent event-event correlation. This procedure enables one to map the seismic data to a growing directed network. There is a unique parameter in this construction, which is the cell size. Since there are no *a priori* rules to determine the cell size, it is important to examine its effects on the structure of the earthquake network. This point was carefully examined in [10]. It was found that the earthquake network has some remarkable properties in its topology. It is a scale-free network, and behaves as a small-world network if directedness of the network is ignored. It was ascertained [10] that these salient features, in fact, remain unaltered by the moderate change of the cell size. Thus, the network approach turned out to highlight complexity of seismicity in a novel manner. Accordingly, researchers have started to examine possible realizations of the earthquake networks by numerical simulations of the models exhibiting self-organized criticality [11].

Here, we discuss the structural and dynamical aspects of the *directed* earthquake network, differently from the small-world picture. In particular, we consider period in the network. This concept is relevant to the question that after how many earthquakes an earthquake returns to the initial location statistically, and therefore it is of obvious interest in view of earthquake prediction.

We define period in a directed network as follows. Take a vertex of the network. There are various closed routes which start from and end at the chosen vertex, in general. Period, $\Pi$, of a chosen closed route is then defined to be the number of edges contained in the route. For example, consider a directed network: $\cdots v_1 \to v_2 \to v_2 \to v_3 \to v_2 \to v_3 \to v_4 \to v_2 \to v_5 \to v_1 \to \cdots$ (see Fig. 1). Being limited to this subgraph, period associated with $v_1$ is 9, whereas $v_2$ has 1, 2, and 3.

We have constructed the directed earthquake networks in California. Firstly, we have divided this district into a collection of cubic cells with two different sizes $10\,\text{km} \times 10\,\text{km} \times 10\,\text{km}$ and $5\,\text{km} \times 5\,\text{km} \times 5\,\text{km}$. Then, the seismic data made available by the Southern California Earthquake Data Center (http://www.scecdc.scec.org/ catalogs.html) has been mapped to growing random networks associated with the two cell sizes by following the procedure explained earlier. The region covered in the data is 29°06.00'N–38°59.76'N latitude and 113°06.00'W–122°55.59'W longitude with the maximal depth 175.99km. The time interval is between 00:25:8.58 on January 1, 1984 and 22:21:52.09 on December 31, 2003. The numbers of independent vertices for the cell sizes $10\,\text{km} \times 10\,\text{km} \times 10\,\text{km}$ and $5\,\text{km} \times 5\,\text{km} \times 5\,\text{km}$ are 3869 and 12913, respectively. The total number of edges (including loops) is 367612, which is equal to the total number of events minus one. We mention that the data used does not include "quarry blasts" but even inclusion of them turned out not to influence the result. In any way, no threshold is put on the value of magnitude.

We have calculated the number of closed routes, $N(\Pi)$, having period, $\Pi$. The results are given in Figs. 2 and 3 for the cell sizes $10\,\text{km} \times 10\,\text{km} \times 10\,\text{km}$ and $5\,\text{km} \times 5\,\text{km} \times 5\,\text{km}$, respectively. Remarkably, they show that $N(\Pi)$ decays as a power law

$$N(\Pi) \sim \Pi^{-\alpha}, \tag{1}$$

where $\alpha$ is a positive exponent. Therefore, there exist a lot of closed routes with significantly long periods in the network, and the law in Eq. (1) makes it difficult to statistically estimate period.

Thus, we have found a scale-invariant law of the period distribution in the directed earthquake network, which indicates that after how many earthquakes an earthquake returns to the initial location. This result manifests the fundamental difficulty in statistically estimating the value of period. As discussed in [10], the distribution of connectivities, $p(k)$, which corresponds to the number of vertices with $k$ edges, also decays as a power law: $p(k) \sim k^{-\gamma}$, showing the scale-free nature of the earthquake network. A phenomenological reason behind this is the following. The seismic data tells us an empirical fact that aftershocks associated with a mainshock tend to return to the locus of the mainshock geographically. The stronger a mainshock is, the more it is followed by aftershocks, yielding a larger value of connectivities. On the other hand, the Gutenberg-Richter law states that cumulative frequency of earthquakes decays as a power law with respect to the value of moment. Such a scale-free property leads to scaling of $p(k)$. A role of a "hub" is played by a mainshock, and the rule of preferential attachment [9] may be realized by aftershocks. Moreover, it is mentioned that the path-length distribution in the directed earthquake network also obeys a power law [12]: $n(l) \sim l^{-\alpha}$, where $n(l)$ is the number of pairs of different vertices in the directed earthquake network, each of which may be linked by the shortest directed routes with $l$ edges. It is quite remarkable that statistics regarding all these topological properties of the earthquake network are scale-invariant. They may highlight criticality/complexity of the earthquake phenomenon and show difficulty of statistical prediction. Further study along this line is expected to give novel insight into seismicity.

The authors were supported by the Grant-in-Aid for Scientific Research of Japan Society for the Promotion of Science.

# Figure Captions

Fig. 1　A schematics description of the earthquake network: $\cdots v_1 \to v_2 \to v_2 \to v_3 \to v_2 \to v_3 \to v_4 \to v_2 \to v_5 \to v_1 \to \cdots$. It has one loop at $v_2$ and one triple edges between $v_2$ and $v_3$.

Fig. 2　(a) The log-log plot of the unnormalized period distribution for the cell size $10\,\text{km} \times 10\,\text{km} \times 10\,\text{km}$. The dots represent the data, whereas the solid line corresponds to the law in Eq. (1) with the exponent $\alpha \cong 1.05$.

(b) The log-log plot of the complementary cumulative number defined by $N(<\Pi) = \int_0^{\Pi} d\Pi' \, N(\Pi')$ associated with $N(\Pi)$ in (a).

All quantities are dimensionless.

Fig. 3　Same analysis as in Fig. 2 for the cell size $5\,\text{km} \times 5\,\text{km} \times 5\,\text{km}$. The value of the exponent is $\alpha \cong 0.96$.

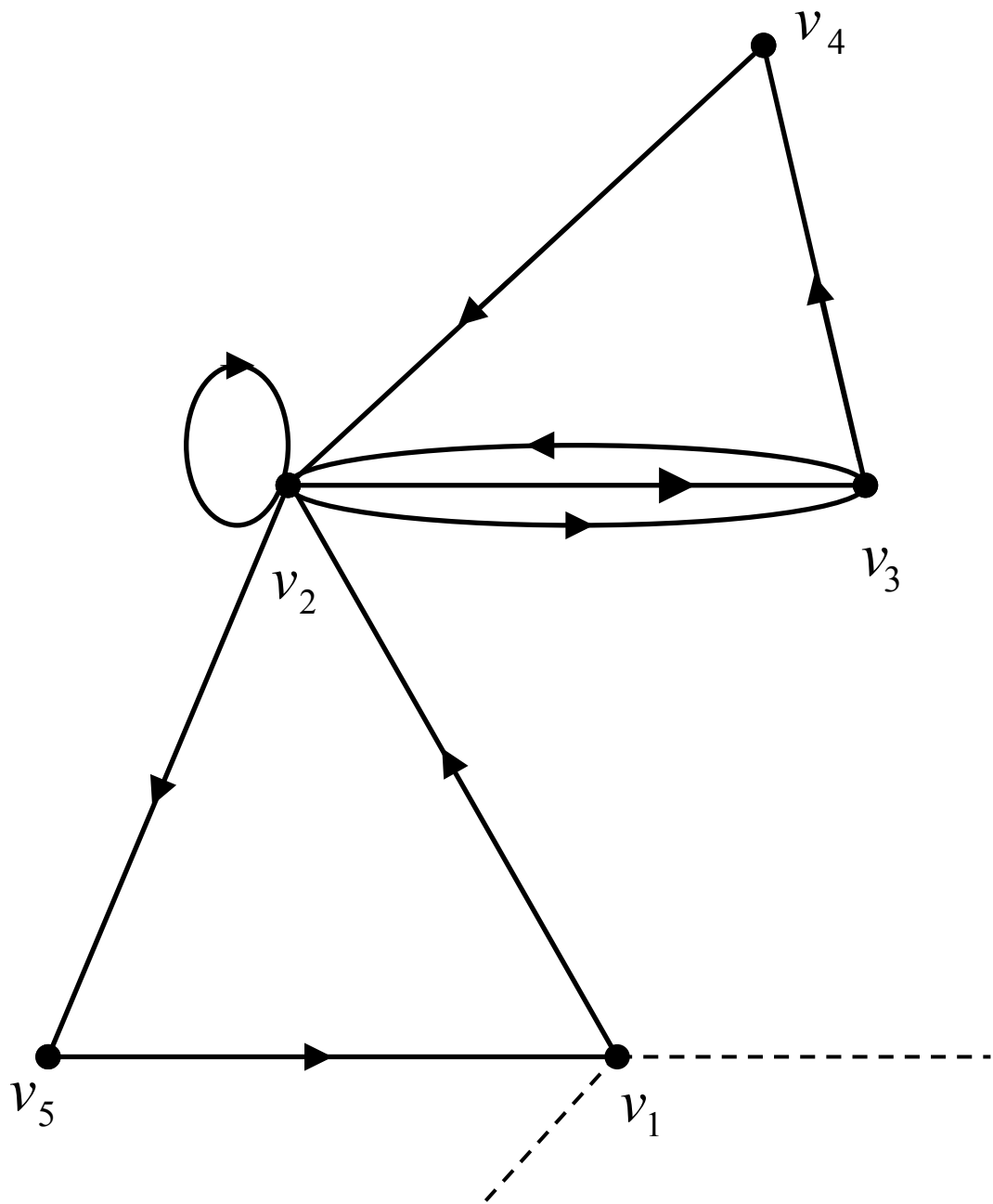

Fig. 1

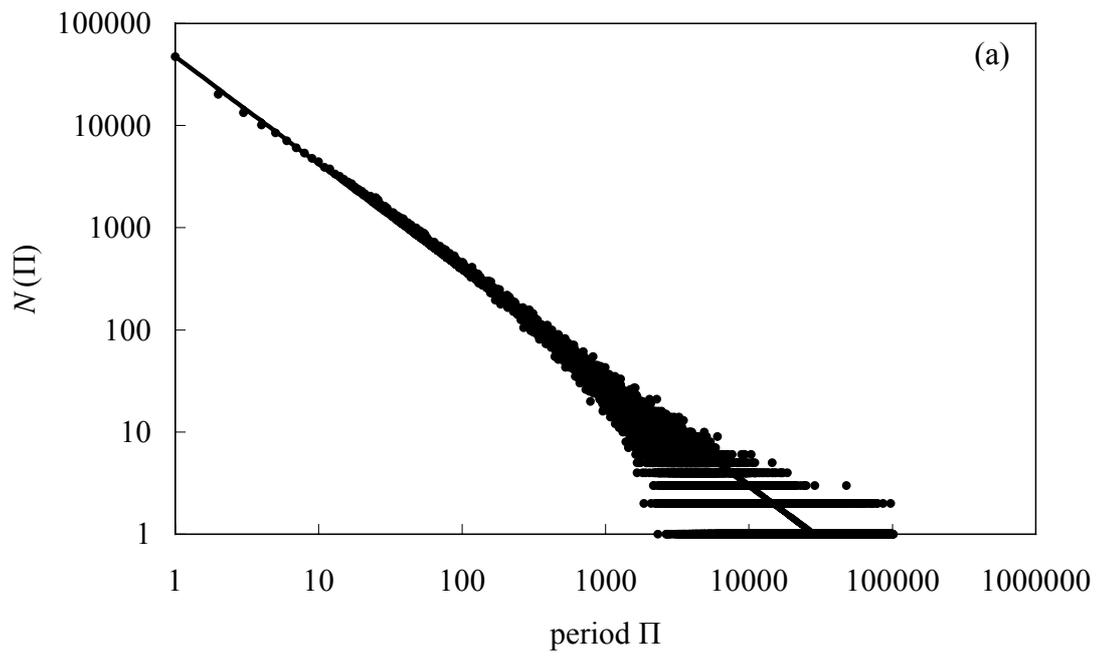

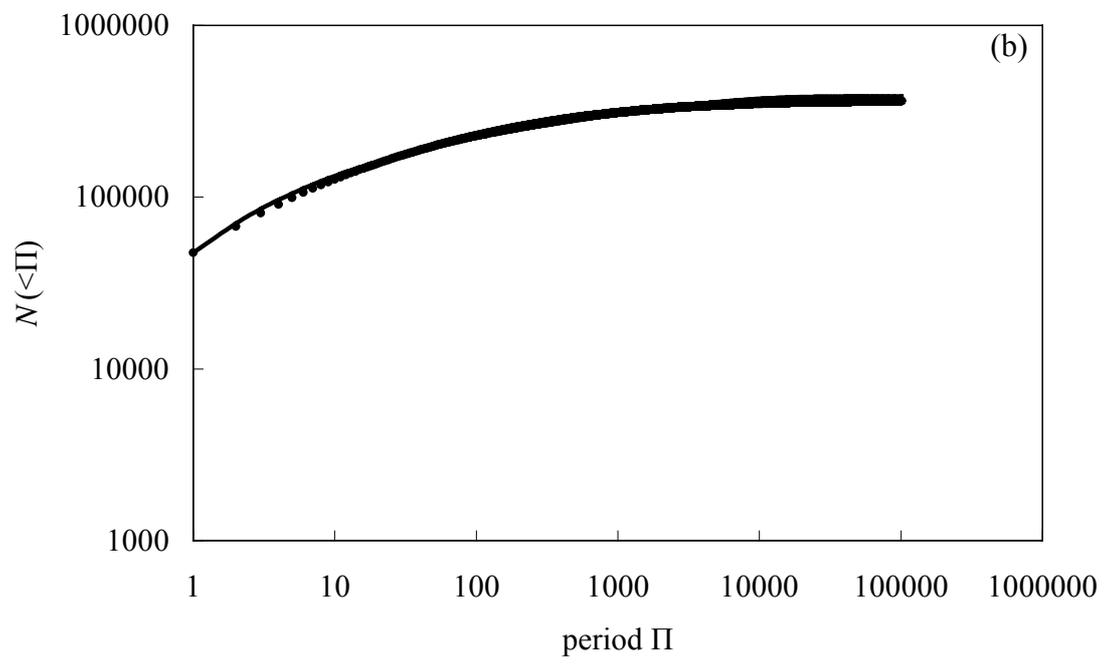

Fig. 2

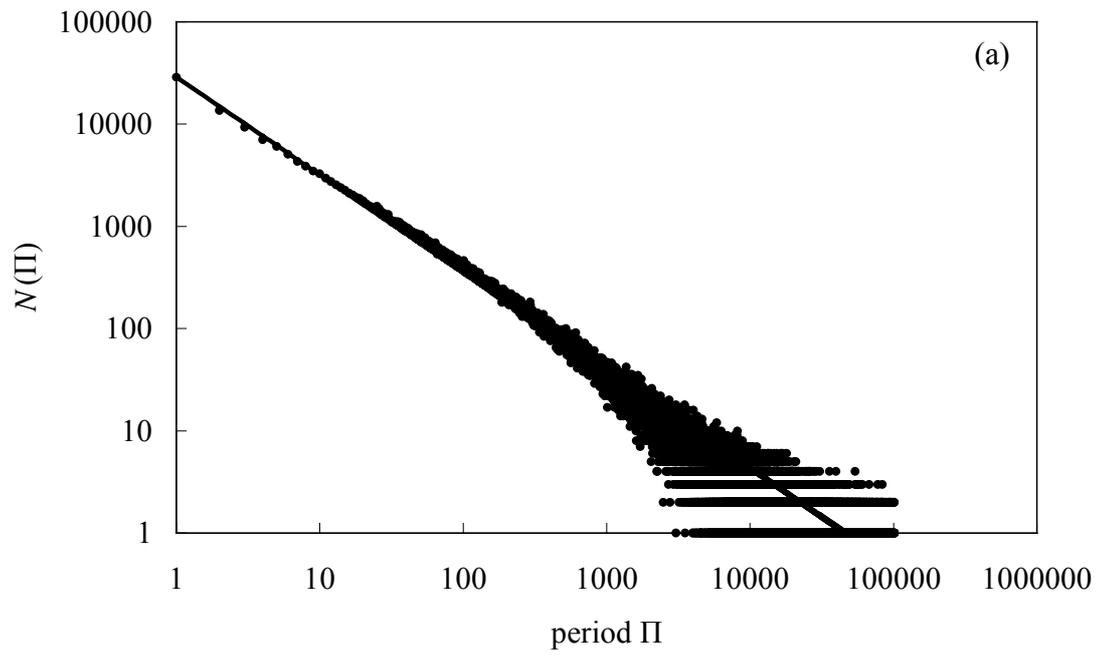

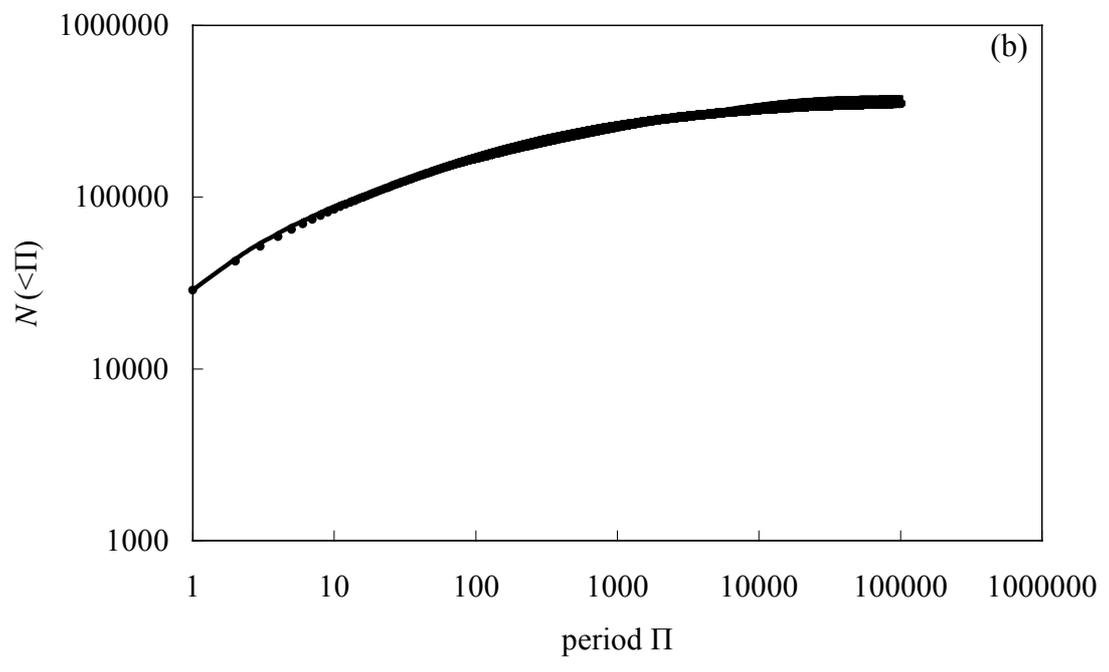

Fig. 3